\documentclass[a4paper,11pt]{article}
\usepackage{pos}
\usepackage{subfigure}

\title{Search for new physics with charm data at BESIII}
\author*[a]{Xinyu Chai} 

 \onbehalf{On behalf of the BESIII Collaboration}

\affiliation[a]{School of Physics, Peking University,\\
  209 Chengfu Road, Beijing, People’s Republic of China}

\emailAdd{chaixy@stu.pku.edu.cn}

\abstract{
Up to now, the Standard Model (SM) remains confronted with several outstanding issues, including the matter–antimatter asymmetry in the universe, as well as the masses, oscillations, and hierarchical structure of neutrinos. These challenges are closely tied to the naturalness and stability of the SM, and have motivated numerous proposals for new physics beyond the SM. The BESIII experiment has accumulated the world’s largest data samples at the 
$J/\psi$ and $\psi(3686)$ resonances, providing an ideal environment for probing signatures of physics beyond the SM. This proceeding presents a summary of recent results published by the BESIII Collaboration on the searches for lepton/baryon number violation (LNV/BNV), charged lepton flavor violation (CLFV), and charmonium weak decays.
}

\FullConference{
$43$rd International Conference on High Energy Physics (ICHEP2026)\\
30 July - 5 August 2026\\
Natal, Brazil\\
}

\begin{document}
\maketitle

\section{Introduction}
The SM has been remarkably successful in describing fundamental particles and their interactions, yet it leaves several outstanding puzzles unresolved, including the matter-antimatter asymmetry of the universe and the origin of neutrino masses. These issues are intimately connected to the naturalness and stability of the SM, and strongly motivate the search for physics beyond the SM.

In the baryon sector, the observed dominance of matter over antimatter necessitates BNV as required by Sakharov's conditions~\cite{Sakharov:1967dj}, which is naturally predicted by grand unified theories and other SM extensions. In the lepton sector, neutrino oscillations indicate non-zero masses, for which the seesaw mechanism~\cite{Georgi:1974sy} provides an explanation via heavy Majorana neutrinos. This framework raises two fundamental questions, the Dirac vs.\ Majorana nature~\cite{Pontecorvo:1957qd} of neutrinos and the existence of heavy Majorana neutrinos, which can be probed through LNV and  CLFV processes. Meanwhile, charmonium weak decays, though kinematically allowed, are highly suppressed in the SM but could be significantly enhanced in many beyond SM scenarios, thus offering a complementary portal for new physics searches~\cite{BESIII:2020nme}.

The BESIII detector, detailed in Ref.~\cite{BESIII:2009fln}, records symmetric $e^+e^-$ collisions provided by the BEPCII storage ring~\cite{Yu:2016cof} in the center-of-mass energy range from 1.85~GeV to 4.95~GeV, with a peak luminosity achieved at $\sqrt{s}=3.773~{\rm GeV}$. With a dataset comprising 10 billion $J/\psi$ events~\cite{BESIII:2021cxx} and 2.7 billion $\psi(3686)$ events~\cite{BESIII:2024lks}, BESIII provides an ideal environment for probing signatures of physics beyond the SM with charm data.

\section{Search for LNV/BNV processes}

\subsection{Study of LNV processes}
%The neutrinoless double beta decay processes are considered as promising way to search for the Majorana neutrinos. The first search for $\Xi^-\rightarrow\Sigma^+e^-e^-$ process~\cite{BESIII:2025ylz} with $\Delta L =2$  is performed using $(10087\pm44)\times10^6$ $J/\psi$ events, where the $\Xi^-$ is selected through $J/\psi \rightarrow\Xi^-\bar{\Xi}^+$ process. Double tag technique and blind analysis strategy are used. The tag side $\bar{\Xi}^+$ is reconstructed from $\bar{\Lambda}\pi^+$ while the $\Xi^-$ is reconstructed using $p\pi^0$. The signal is extracted by fitting $M_{p\pi^0}$, which is shown in Fig.~\ref{fig:mgg}, and no evidence of significant signal yield is observed in this invariant mass spectrum. By using Bayesian approach, the upper limit (UL) is set to be $\mathcal{B}(\Xi^-\rightarrow\Sigma^+e^-e^-)$ at $90\%$ confidence level (C.L.).

%And there are also some first searches for the meson neutrinoless double beta decay processes. For $\eta\rightarrow\pi^+\pi^+e^-e^-$ process, its UL of BF is set to be $4.6\times10^{-6}$. For $\omega\rightarrow\pi^+\pi^+e^-e^-$ process, its UL of BF is set to be $2.8\times10^{-6}$.   For $\phi\rightarrow\pi^+\pi^+e^-e^-$ process, its UL of BF is set to be $3.7\times10^{-6}$.  And for the $J/\psi\rightarrow K^+K^+e^-e^-$ process, its UL of BF is set to be $2.1\times10^{-9}$. And the spectrum used to extracted signal yields are shown in Fig.~\ref{}.

\begin{figure}[!htbp]
\centering
\includegraphics[width = 0.45\textwidth]{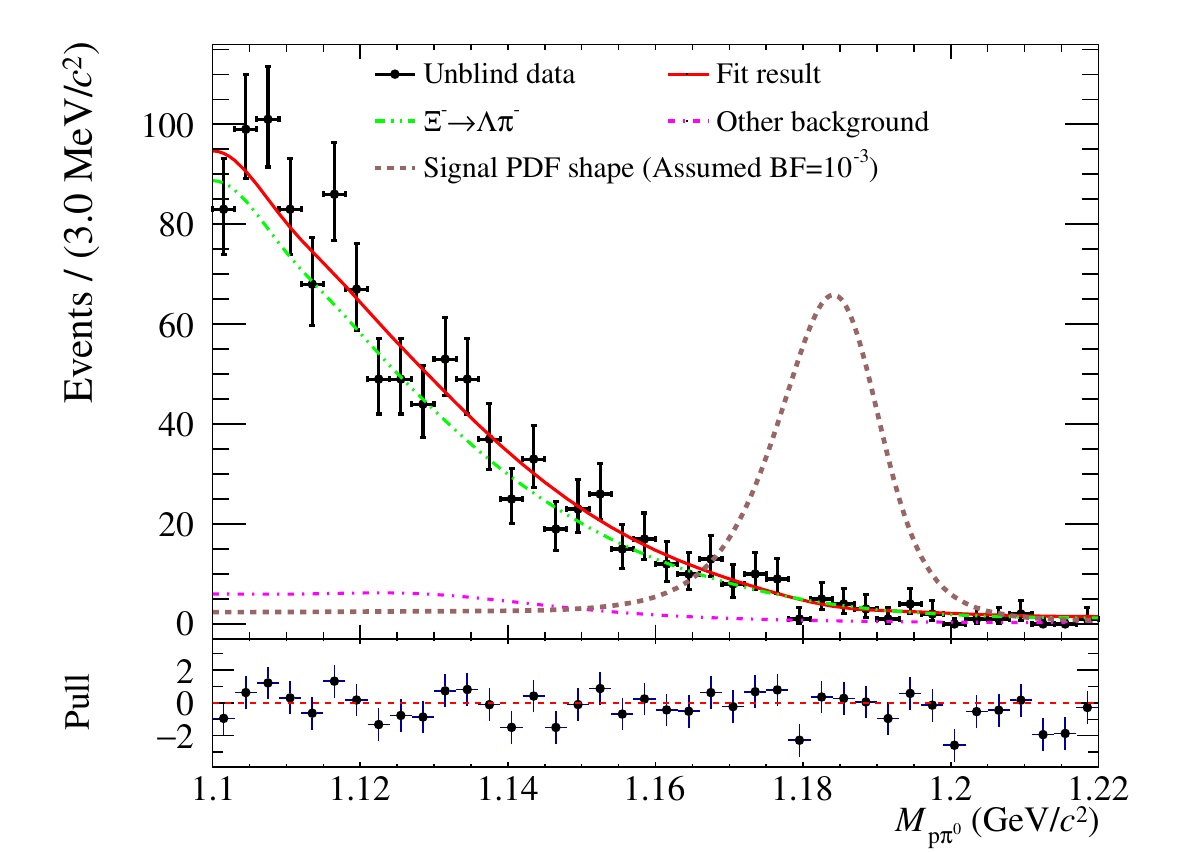}
\caption{The upper panel shows fit result on the $M_{p\pi^0}$ distribution. The black dots with error bars correspond to data samples, the red solid line indicates the fitting result, the green dashed-dotted line corresponds to the component of the $\Xi^-\rightarrow\Lambda\pi^-$ background, the brown dotted line represents the signal PDF shape and the pink dashed-dotted line illustrates the component of other backgrounds. The lower panel shows the pull distribution defined as $\frac{y_{\rm data}-y_{\rm fit}}{\sigma}$, where the $\sigma$ considers the statistical and systematic uncertainties.}
\label{fig:mgg}
\end{figure}

The neutrinoless double beta decay processes are considered promising channels for probing the Majorana nature of neutrinos. A first search for the $\Delta L = 2$ process $\Xi^- \rightarrow \Sigma^+ e^- e^-$ has been performed using $(10087 \pm 44) \times 10^6$ $J/\psi$ events, where the $\Xi^-$ candidates are produced via the $J/\psi \rightarrow \Xi^- \bar{\Xi}^+$ decay \cite{BESIII:2025ylz}. The analysis employs a double-tag technique and a blind analysis strategy. The tag-side $\bar{\Xi}^+$ is reconstructed via $\bar{\Lambda} \pi^+$, while the signal-side $\Xi^-$ is reconstructed using the $p\pi^0$ final state. The signal yield is extracted from a fit to the $M_{p\pi^0}$ distribution, shown in Fig.~\ref{fig:mgg}, and no significant signal excess is observed. An upper limit (UL) on the branching fraction (BF) $\mathcal{B}(\Xi^- \rightarrow \Sigma^+ e^- e^-)$ is setto be $2.0\times10^{-5}$ at the $90\%$ confidence level using a Bayesian approach.

In addition to the hyperon decay, first searches for neutrinoless double beta decay have also been extended to meson systems. For the processes $\eta \rightarrow \pi^+ \pi^+ e^- e^-$~\cite{BESIII:2025ycw}, $\omega \rightarrow \pi^+ \pi^+ e^- e^-$~\cite{BESIII:2025jyk}, $\phi \rightarrow \pi^+ \pi^+ e^- e^-$~\cite{BESIII:2023pqp}, and $J/\psi \rightarrow K^+ K^+ e^- e^-$~\cite{BESIII:2025wtj}, the corresponding ULs on their BFs are set to $4.6 \times 10^{-6}$, $2.8 \times 10^{-6}$, $3.7 \times 10^{-6}$, and $2.1 \times 10^{-9}$, respectively. 
%The invariant mass spectra used for signal extraction are displayed in Fig.~\ref{fig:lnv}, which shows very low background level.

\subsection{Study of BNV processes}
Using $(10087 \pm 44) \times 10^6$ $J/\psi$ events, a first search for $\Lambda$-$\bar{\Lambda}$ oscillation~\cite{BESIII:2024gcd} is performed via the $J/\psi \rightarrow \Lambda \bar{\Lambda}$ decay. Events in which a $\Lambda$ ($\bar{\Lambda}$) oscillates into a $\bar{\Lambda}$ ($\Lambda$) are classified as wrong-sign (WS) events, while those without oscillation are classified as right-sign (RS) events. The UL on the time-integrated oscillation rate is calculated as
\begin{equation}
    P(\Lambda)=\frac{\int_0^{\infty} \sin^2(\delta m_{\Lambda\bar{\Lambda}} \cdot t) \, e^{-t/\tau_{\Lambda}} \, dt}{\int_0^{\infty} e^{-t/\tau_{\Lambda}} \, dt}
    =\frac{\mathcal{B}(J/\psi \rightarrow \Lambda\Lambda + c.c.)}{\mathcal{B}(J/\psi \rightarrow \Lambda\bar{\Lambda})}
    < \frac{N_{\rm WS}^{\rm UL}}{N_{\rm RS}^{\rm obs} / \epsilon_{\rm RS}}
    = 1.4 \times 10^{-6},
\end{equation}
where the $\tau_{\Lambda}$ is the mean lifetime of $\Lambda$, and the RS signal yield is $N_{\rm RS}^{\rm obs} = 3123264 \pm 1767$ and the detection efficiencies for the RS is determined to be $\epsilon_{\rm RS}=34.3\%$. In the WS search, no signal or background events are observed, and the UL on the number of WS events is determined to be $N_{\rm WS}^{\rm UL} = 12.0$ at the $90\%$ confidence level. Correspondingly, the UL on the oscillation parameter is set to $\delta m_{\Lambda\bar{\Lambda}} < 2.1 \times 10^{-18}~{\rm GeV}$, which translates to a lower limit on the oscillation time of $\tau_{\rm osc} > 3.1 \times 10^{-7}~{\rm s}$. This result imposes a stricter constraint than the previous measurement based on $J/\psi \rightarrow pK^-\bar{\Lambda}$ decays~\cite{BESIII:2023tge}.

\subsection{Study of LNV\& BNV process}
\begin{figure}[!htbp]
\centering
\includegraphics[width = 0.45\textwidth]{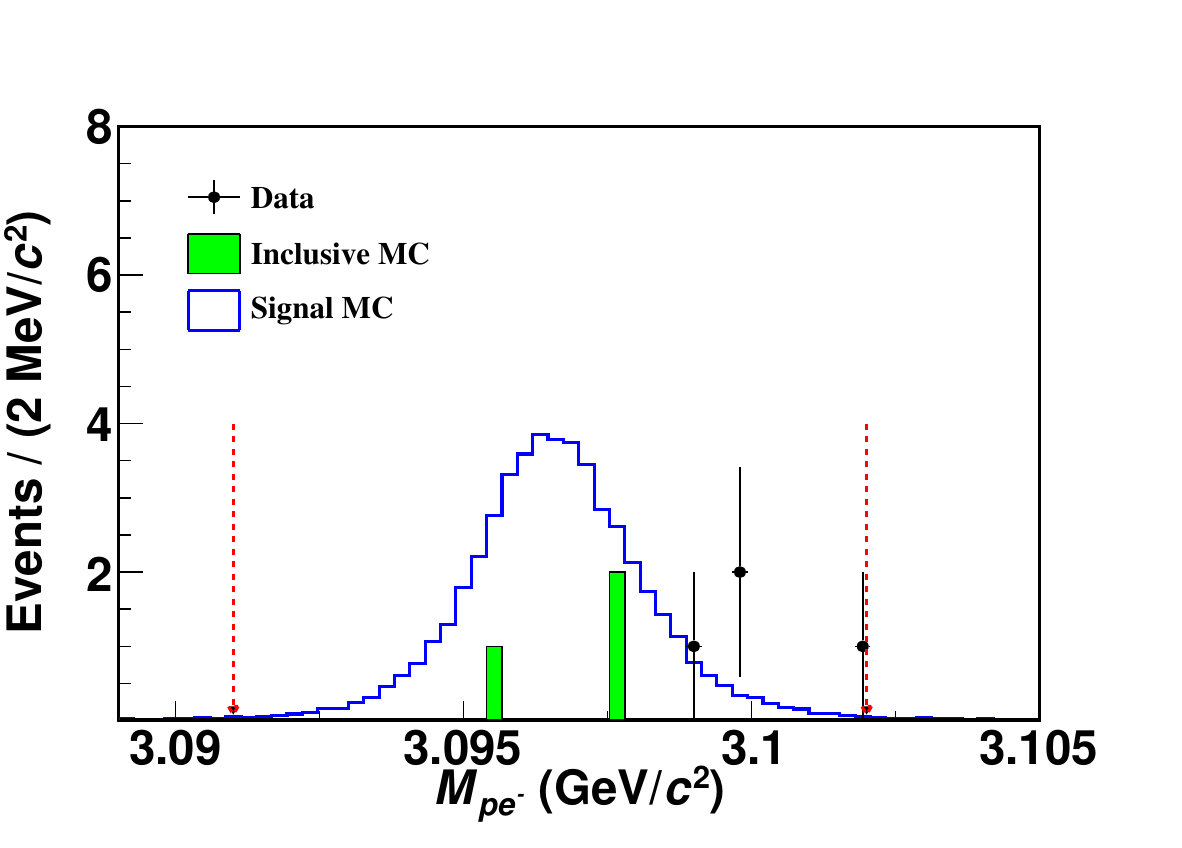}
\caption{The $M_{pe^-}$ distribution shows data points with error
bars, the background as a green histogram, the signal MC
as a blue histogram, and the signal region is marked by red
dashed arrows.}
\label{fig:lnvbnv}
\end{figure}

The first search for the $J/\psi \rightarrow pe^- +c.c.$~\cite{BESIII:2025jtv} has been performed via $\psi(3686)\rightarrow \pi^+\pi^-J/\psi$ based on $(2712.4 \pm 14.3) \times 10^6$ $\psi(3686)$ events. As shown in Figure.~\ref{fig:lnvbnv}, no obvious signals have been observed. By using the frequentist method~\cite{Rolke:2004mj}, the UL on the BF at 90\% C.L. is set to be $\mathcal{B}(J/\psi\rightarrow pe^- +c.c.)<3.1\times10^{-8}$.

\section{Search for CLFV processes}
The observation of neutrino oscillations establishes that lepton flavor is not an exact symmetry in nature, implying that CLFV processes can occur in the SM extended with neutrino masses. Searches for CLFV have been conducted in both vector and pseudoscalar quarkonium states.

\begin{figure}[!htbp]
\centering
\subfigure[$\psi(3686)\rightarrow e\mu$]{
\centering
\includegraphics[width=0.35\textwidth]{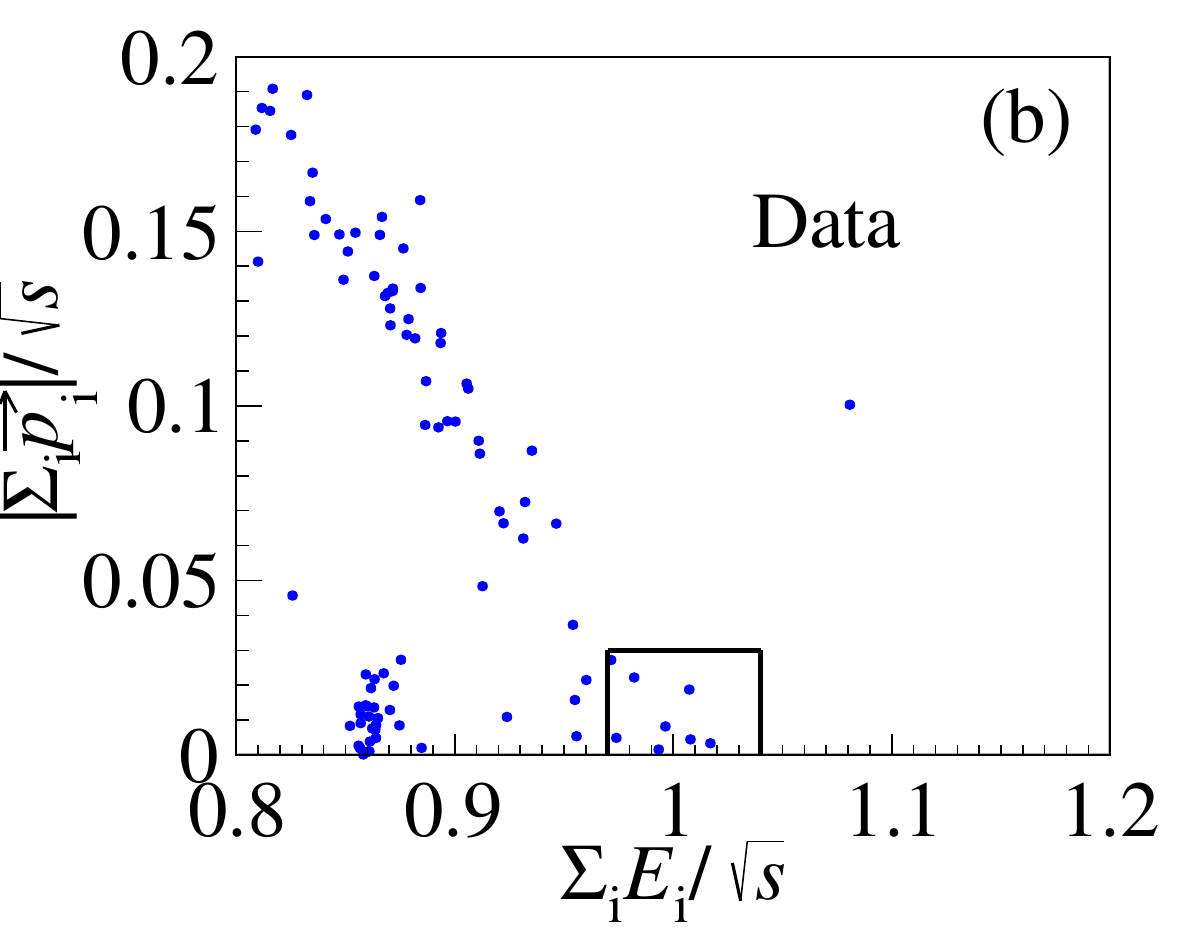}
}%
\subfigure[$\eta\rightarrow e\mu$]{
\centering
\includegraphics[width=0.35\textwidth]{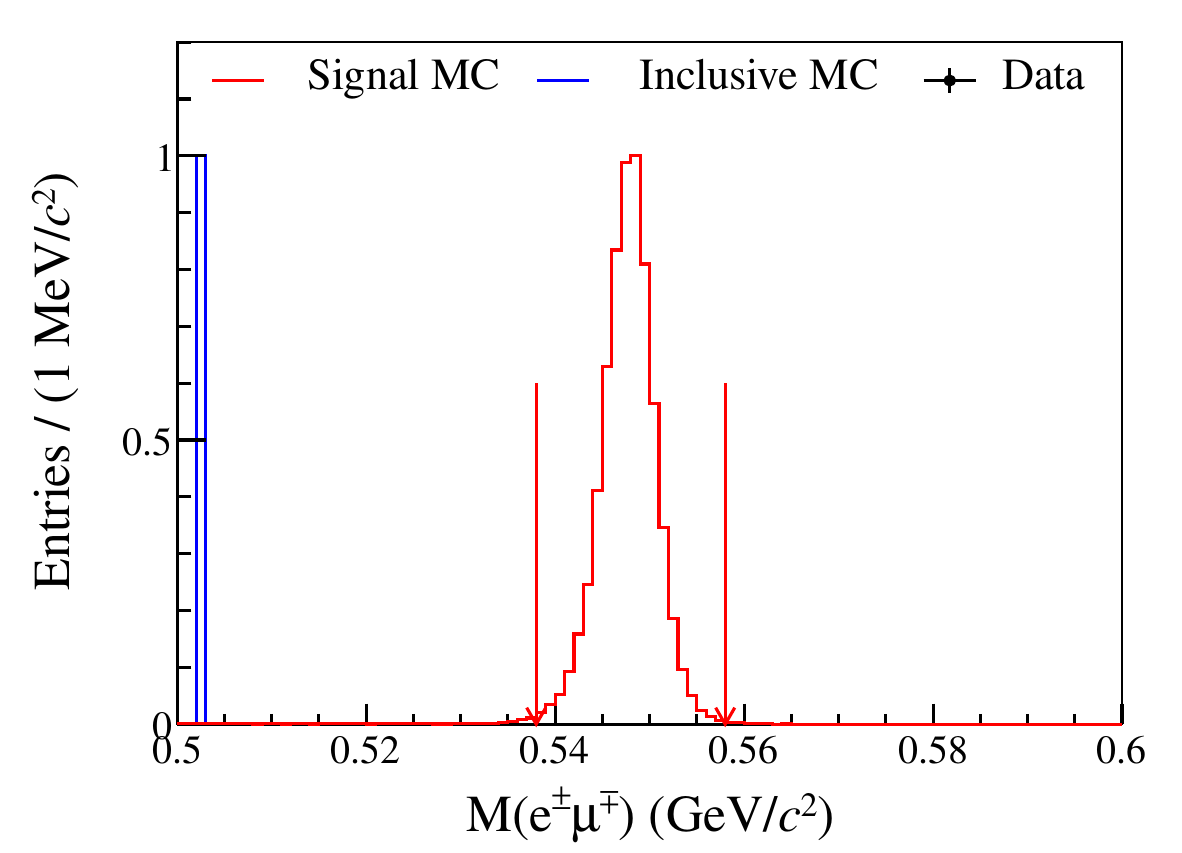}
}%
\subfigure[$\eta\prime\rightarrow e\mu$]{
\centering
\includegraphics[width=0.35\textwidth]{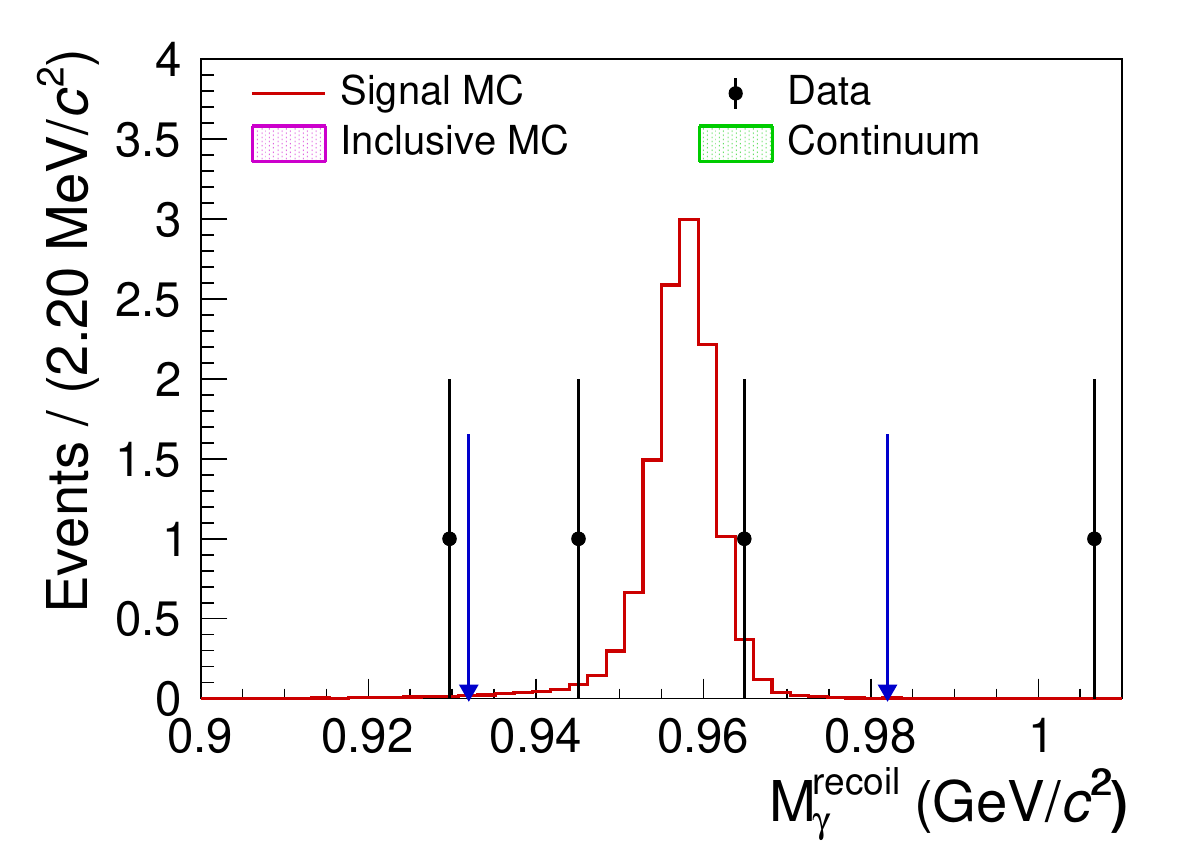}
}%

\caption{(a) The distribution of $|\Sigma_i\vec{p}_i|/\sqrt{s}$ versus $|\Sigma_i\vec{E}_i|/\sqrt{s}$ for $\psi(3686)\rightarrow e\mu$ process;
(b) The $M_{e\mu}$ distribution for $\eta\rightarrow e\mu$ process;
(c) The $M^{\rm recoil}_{\gamma}$ distribution of candidate events for $\eta\prime\rightarrow e\mu$ process.}
\label{fig:clfv}
\end{figure}

For the vector state $\psi(3686)$, a search for the decay $\psi(3686) \rightarrow e\mu$ has been performed using $(2367.0 \pm 11.1) \times 10^6$ events \cite{BESIII:2025emu}. As shown in Fig.~\ref{fig:clfv}(a), eight candidate events are observed in the signal region, while the expected background is estimated to be $6.2 \pm 1.4$, indicating no significant excess. An UL on the BF is set using the profile likelihood method \cite{Rolke:2004mj} to be $\mathcal{B}(\psi(3686) \rightarrow e\mu) < 1.4 \times 10^{-8}$ at the $90\%$ confidence level.

For the pseudoscalar states $\eta$ and $\eta^\prime$, the searches are performed using the $J/\psi \rightarrow \gamma \eta^\prime$ process, where the $\eta$ is accessed via the subsequent decay $\eta^\prime \rightarrow \pi\pi \eta$.
The invariant mass spectra used for signal extraction in these two channels are displayed in Fig.~\ref{fig:clfv}(b) and (c), respectively. 
For the $\eta^\prime \rightarrow e\mu$ process \cite{BESIII:2026sgd}, the UL on the BF is set to $\mathcal{B}(\eta^\prime \rightarrow e\mu) < 6.3 \times 10^{-7}$ at $90\%$ C.L., which is three orders of magnitude stricter than the previous limit from CLEO \cite{CLEO:1999nsy}. For the $\eta \rightarrow e\mu$ process \cite{BESIII:2026wbm}, the UL is set to $\mathcal{B}(\eta \rightarrow e\mu) < 6.8 \times 10^{-7}$ at $90\%$ C.L., improving by one order of magnitude over the limit set by the $pd$ scattering experiment in 1996 \cite{White:1995jc}. The corresponding Wilson coefficient constraints for these three channels are also provided in Refs.~\cite{BESIII:2025emu, BESIII:2026sgd, BESIII:2026wbm}.

\section{Search for Charmonium weak decay processes}
Within the SM, the inclusive BFs of rare weak decays of the $J/\psi$ involving one charmed meson are predicted to be of order $10^{-8}$ or lower. However, these processes can be significantly enhanced in various new physics scenarios, such as the top-color model~\cite{Hill:1994hp}, the minimal supersymmetric standard model with or without R-parity violation~\cite{Aulakh:1982yn}, and the two Higgs doublet model~\cite{Glashow:1976nt}. Consequently, such charmonium weak decays serve as promising probes for new physics beyond the SM. To date, BESIII has performed searches for these decays, and no significant signal is observed in any of the channels.

For the $c \rightarrow u$ flavor-changing neutral current (FCNC) weak decays, the ULs on the BFs at $90\%$ C.L. are set to $1.4 \times 10^{-7}$ for $J/\psi \rightarrow D^0 K^{*0} + c.c.$~\cite{BESIII:2025hqb}, $9.1 \times 10^{-8}$ for $J/\psi \rightarrow \gamma D^0 + c.c.$~\cite{BESIII:2024kkx}, and $3.1 \times 10^{-8}$ for $J/\psi \rightarrow D^0 \mu^+ \mu^- + c.c.$~\cite{BESIII:2025wsu}, respectively. For the $c \rightarrow d$ weak decays, the corresponding ULs at $90\%$ C.L. are $8.0 \times 10^{-7}$ for $J/\psi \rightarrow D_s^- \rho^+ + c.c.$, $4.1 \times 10^{-7}$ for $J/\psi \rightarrow D_s^- \pi^+ + c.c.$~\cite{BESIII:2025rjn}, and $9.9 \times 10^{-8}$ for $J/\psi \rightarrow D_s^- e^+ \nu_e + c.c.$~\cite{BESIII:2025frp}, respectively.

\section{Summary}

%To search for the new physics beyond the SM, BESIII has performed lots of first or updated study on the LNV, BNV, CLFV and charmonium weak decays with the large data charm data sample, which provide important experimental constraints.  No significant signal is found in these scenarios, and the exclusion limits are helpful to constrain the new physics models beyond the SM. As more data becomes available, we are expecting more definitive results coming in the future.

To probe for physics beyond the SM, BESIII has performed a series of first searches and updated studies on LNV, BNV, CLFV, and charmonium weak decays, utilizing the world's largest samples of charmonium events together with high-statistics charm data. These measurements provide important experimental constraints on a wide range of new physics scenarios. Although no significant signal has been observed in any of the channels examined, the stringent upper limits established in these searches offer valuable guidance for constraining the parameter spaces of various beyond-SM models. With the continued accumulation of data at BEPCII, further improvements in sensitivity are expected, which will either reveal evidence of new physics or push the current limits to even more stringent levels.

\end{document}